\documentclass[structabstract]{aa}  

\usepackage{graphicx}
\usepackage{color}
\usepackage{natbib}
\bibpunct{(}{)}{;}{a}{}{,} %
\usepackage{txfonts}
\def\aap{A\&A}

\begin{document}

   \title{AMBER and CRIRES observations of the binary sgB[e] star HD~327083: evidence of a gaseous disc traced by CO bandhead emission\thanks{Based on data obtained at the European Southern Observatory, Paranal, Chile via the observing programmes 385.D-0513 and 087.D-0889.}}

\author{H.E. Wheelwright
\inst{1}
\and
W.J. de Wit\inst{2}	
\and{G. Weigelt}\inst{1}
\and
R.D. Oudmaijer\inst{3}
\and
J.D. Ilee\inst{3}
}

   \institute{Max-Planck-Institut f\"{u}r Radioastronomie, Auf dem H\"{u}gel 69,
53121 Bonn, Germany\\\email{hwheelwright@mpifr-bonn.mpg.de}
\and
European Southern Observatory, Alonso de Cordova 3107, Vitacura, Santiago, Chile
\and
School of Physics and Astronomy, University of Leeds, Leeds LS2 9JT, UK
}

   \date{Received Month dd, yyyy; accepted Month dd, yyyy}
\abstract
{\object{HD 327083} is a supergiant B[e] star that forms a binary
  system with an orbital semi-major axis of approximately 1.7~AU.}
{Our previous observations using the VLTI and AMBER in the medium
  resolution $K-$band mode spatially resolved the environment of HD
  327083. The continuum visibilities obtained indicate the presence of
  a circumbinary disc. CO bandhead emission was also
  observed. However, due to the limited spectral resolution of the previous
  observations, the kinematic structure of the emitting material could
  not be constrained. In this paper, we address this and probe the source of
  the CO emission with high spectral resolution and spatial precision.}
{To determine the properties and kinematics of its CO emitting region,
  we have observed HD~327083 with high spectral resolution (25 \&
  6~$\mathrm{km\,s^{-1}}$) using AMBER and CRIRES. The observations
  are compared to kinematical models to constrain the source of the
  emission.}
{The multi-epoch AMBER spectra obtained over 5~months contain no
  evidence that the CO $\mathrm{1^{st}}$ overtone emission of
  HD~327083 is variable. This indicates that the structure of the
  emitting region is not strongly dependent on orbital phase. It is
  shown that the CO bandhead emission can be reproduced using a model
  of a Keplerian disc with an inclination and size consistent with our
  previous VLTI observations. The model is compared to AMBER
  differential phase measurements, which have a precision as high as
  $\sim$30~$\mu$as. A differential phase signal corresponding to
  0.15~mas ($\sim$5~$\sigma$) is seen over the bandhead emission,
  which is in excellent agreement with the model that fits the CRIRES
  observations. In comparison, a model of an equatorial outflow, as
  envisaged in the standard sgB[e] scenario, does not reproduce the
  observations well.}
{We present a direct test of the circumstellar kinematics of the
  binary sgB[e] star HD 327083 using both spatial and spectral
  information. The excellent agreement between the disc model and
  observations in the spatial and spectral domains is compelling
  evidence that the CO bandhead emission of HD~327083 originates in a
  circumbinary Keplerian disc. In contrast, the model of an equatorial
  outflow cannot reproduce the observations well. This suggests that
  the standard sgB[e] scenario is not applicable to HD~327083, which
  supports the hypothesis that the B[e] behaviour of HD~327083 is due
  to binarity.}

   \keywords{Stars: circumstellar material --
	     Stars: early type --
	     Stars: emission-line, Be --
	     Stars: mass-loss--
             Stars: individual: HD 327083}
\titlerunning{Tracing the origin of the CO bandhead emission of HD~327083}
\authorrunning{H.E. Wheelwright et al.}
   \maketitle
%

\section{Introduction}

Massive stars play a crucial role in multiple areas of
astrophysics. As a result, a complete understanding of many
astrophysical phenomena requires an understanding of the evolution of
massive stars. In turn, this requires knowledge of how massive stars
lose mass \citep[see e.g.][]{Puls2008}. Supergiant B[e] {{(sgB[e])}}
stars are important objects in this regard as they are massive objects
in a late evolutionary stage which also exhibit signs of enhanced mass
loss \citep[see e.g.][]{Lamers1998}. In addition, it has been
suggested that there could be an evolutionary link between sgB[e]
stars and the luminous blue variables \citep[LBVs,][]{Zickgraf2006}. In turn,
LBVs are generally thought to be the precursors of
the Wolf-Rayet stars; the final phase in massive star evolution
\citep[see e.g.][]{Jorick_lbv}. Therefore, the asymmetric geometries
that are required to produce phenomena such as gamma ray bursts may be
constructed in the sgB[e] stage. Consequently, it is important to
study the circumstellar geometry of sgB[e] stars and determine what is
responsible for their mass loss.

\smallskip

With this in mind, we recently studied the sgB[e] star HD~327083 with
the VLTI and AMBER in the medium resolution $K-$band mode
\citep[see][hereafter W12]{Me_HD_327083_1}. High spatial resolution
observations were required to probe the geometry of the object's
circumstellar environment on milli-arcsecond scales. In turn, this was
necessary to elucidate the ``bracket'' behaviour of HD~327083, for
which there have been several hypotheses. On the basis of a comparison
between optical spectroscopy and a NLTE model of an expanding
atmosphere, \citet{Machado2003} suggest that the object may be close
to the {{LBV phase}}. This would make it a key object to study the
link, if any, between the sgB[e] and LBVs. However, there is an
alternative hypothesis. \citet[][hereafter M2003]{Miro2003} detected
an unresolved binary companion via radial velocity variations. Based
on the relatively short period of the variations, $\sim$6~months,
these authors suggest that the system is close enough to interact. In
this case, the material traced by the object's infrared excess may be
mass lost as the result of binary interactions, as proposed for other
objects showing the B[e] phenomenon
\citep[see][]{Millour2009,Millour2011,Kraus_v921_2012}. 

\smallskip 

We now briefly summarise the results from W12. Our observations with
the VLTI and AMBER spatially resolved the material responsible for the
$K-$band excess of HD~327083. Using simple geometrical models, we
found that the continuum visibilities could be reproduced by an
elongated ring with a central radius of approximately 6.6~AU. The
location of this ring is consistent with the expected dust sublimation
radius of a star with the stellar parameters reported by M2003;
$6<R_{\rm{sub}}<25$~AU using the standard equation
\citep[see][]{Monnier2002}. A non-zero closure phase was observed over
the continuum emission and this was attributed to the binary
companion. The closure phase suggested that the companion should be
located within the elongated ring. Therefore, the geometrical model
was associated with the inner rim of a dusty, circumbinary disc. Such
a configuration is more reminiscent of mass lost as the result of
binary interactions \citep[see e.g.][]{Millour2011} than intrinsic
mass loss via an equatorial outflow, as depicted in the standard
sgB[e] scenario \citep[see e.g.][]{Zickgraf1985}.  Therefore, the
observations to date favour the interacting binary hypothesis.

\smallskip

Probing the mass loss process of the system further requires
constraining the kinematics of the circumbinary material and the
conditions that give rise to the occurrence of the B[e]
phenomenon.  Such constraints can be determined from spectrally
resolved observations of CO bandhead emission. Therefore, we obtained
new high spectral resolution (25 \& 6~$\mathrm{km\,s^{-1}}$) observations
of HD~327083's CO bandhead emission with AMBER and CRIRES. Here we
present these observations, which provide sub-milli-arcsecond
precision constraints on the geometry of the emitting material. The CO
bandhead emission of HD~327083, initially observed by
\citet{McGregor1988}, is spectrally resolved for the first time and we
observe a differential phase signature over the emission. This paper
presents the observations and modelling of the CO emitting material,
and it is structured as follows. In Sect. \ref{obs_and_data} we
describe the observations and subsequent data reduction. We present
the observational results in Sect. \ref{res}. These are then analysed
and modelled in Sect. \ref{ana} and the implications are discussed in
Sect. \ref{disc}. The conclusions are presented in Sect. \ref{conc}.

\section{Observations and data reduction}

\label{obs_and_data}


\subsection{CRIRES}
HD~327083 was observed with the near-IR cryogenic high resolution
spectrograph CRIRES of the VLT \citep{CRIRES} during
the night of June 28, 2010. The target was included in our survey of
sixteen Galactic, $K-$band bright sgB[e] candidates and unclassified B[e]
stars. The aim of the project is to study in detail the
kinematic signature of the CO first overtone bandhead of B[e] stars
in general (Muratore et al. in prep.). Here we make use of the data
for this object. Observations were executed with a slit-width of
0.4\arcsec$\,$ delivering a spectral resolution of approximately 50,000
(or $\rm 6\,km\,s^{-1}$). The data were obtained applying a
standard nodding on-slit strategy to remove sky and detector glow,
without AO correction and applying a small jittering offset at each
nod position to better remove bad pixels. The DIT was chosen bearing
in mind the non-linearity regime of the CRIRES InSb Aladdin III
detectors. A telluric standard was taken immediately after the science
targets, close in airmass. The seeing conditions were poor as the
seeing varied between 1.75 and 2.25\arcsec. However, the brightness of
the target, $K=$3.3~mag., ensured the data are still of high quality. Data
reduction was conducted in a standard fashion utilising the ESO
provided CRIRES pipeline.

\subsection{AMBER}

\begin{center}
  \begin{figure}
    \begin{center}
      \includegraphics[width=0.45\textwidth]{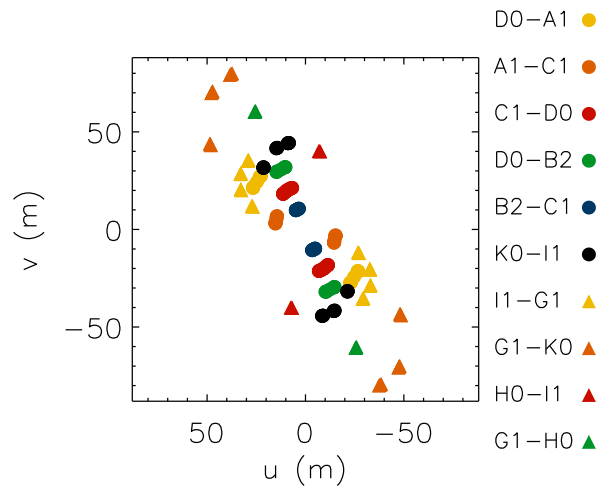}
      \caption{The projected baselines of the AMBER observations of HD~327083.\label{FIG:uv_cov}}
    \end{center}
  \end{figure}
\end{center}

HD~327083 was observed with the VLTI and the 3-beam combiner
instrument AMBER \citep{Petrov2007} in the high spectral resolution
$K-$band mode centred on 2.288~$\mu$m. This setting delivers a
spectral resolution of $R=$12,000 and a wavelength range of
2.265 to 2.311~$\mu$m. FINITO was used to provide fringe
tracking. Observations were conducted on 6 occasions between May
$\mathrm{2^{nd}}$ 2011 and September $\mathrm{22^{nd}}$ 2011. The
observations were carried out employing the 1.8m Auxiliary Telescopes
(ATs). A log of the observations is presented in Table
\ref{TABLE:amber_obs} and the projected baselines are presented in
Fig. \ref{FIG:uv_cov}. In general, short baselines were used as our
previous observations demonstrate that this object is relatively
extended. The seeing conditions were typically good (the mean seeing
was $\sim$0.75\arcsec), which was important to facilitate these high
spectral resolution observations. Observations of HD~327083 took place
between observations of calibrator objects (HD~151078 and HD~161068).

\begin{center}
  \begin{table*}
    \begin{center}
      \caption{The log of the AMBER observations.\label{TABLE:amber_obs}}
      \begin{tabular}{l l c c l l l}
        \hline
        Date & Stations & Seeing & Coherence time & DIT & Baseline&PA \\
             &          & (\arcsec) & (ms) & (s) & (m) & ($^{\circ}$)\\
        \hline
        \hline
        02-05-2011 & H0-I1-G1 & 0.46 & 5.3&12.0 & 40.7/38.6/65.7 & 169.9/58.1/23.0\\
        07-05-2011 & D0-A1-C1 & 0.68 & 4.1&12.0 & 34.7/15.7/21.8 & 45.4/71.7/26.9\\
        12-06-2011 & D0-B2-C1 & 0.88 & 4.6&12.0 & 33.2/11.0/22.1 & 22.2/22.1/22.2\\
        27-07-2011 & K0-I1-G1 & 1.03 & 2.1&12.0& 44.2/43.6/84.9 & 19.3/49.0/34.0 \\
        04-08-2011 & K0-I1-G1 & 0.74 & 2.1&12.0& 45.1/45.9/88.1 & 11.1/39.5/25.4 \\
        22-09-2011 & K0-I1-G1 & 0.77 & 3.1&12.0 & 38.2/29.5/65.1 & 33.9/66.2/47.9\\
        \hline
      \end{tabular}
      \tablefoot{The baseline lengths and position angles (PAs)
        represent time-averaged values for a given night.}
    \end{center}
  \end{table*}
\end{center}

The AMBER observations were reduced in a standard fashion using the
amdlib software version 3.0.3
\citep[see][]{Tat-amdlib,Chelli2009}. The processes involved,
i.e. determining the pixel-to-visibility-matrix and converting the
observed fringe patterns to measurements of the coherent flux, are
standard operations with this software and are not described
here. Once the interferometric observables are calculated, the general
approach is to subject each block of frames to certain selection
criteria before averaging the selected frames into a single
measurement. Since we employed the high spectral resolution mode of
AMBER with a long integration, and the data exhibit a range of frame
signal-to-noise-ratios, it is challenging to accurately calibrate the
visibilities. Therefore, we do not make use of absolute visibilities
in this paper. As we are not concerned with temporal variations in the
transfer function, we merge all consecutive observations before
performing frame selection and averaging. This provides a higher
resultant signal-to-noise-ratio (SNR) \citep[see e.g.][]{Stefl2009}. A
high frame selection rate, 80 per cent, was used as this generally
provides the highest quality phase information and our analysis in
this paper concentrates on these measurements.

\smallskip

Wavelength calibration of the AMBER data was conducted using the
telluric lines present in the spectra. To provide a reference
wavelength for the telluric lines observed, we utilised a high
resolution ($R=$40,000) spectrum of the Earth's telluric features made
available by the NSO/Kitt Peak
Observatory\footnote{http://www.eso.org/sci/facilities/paranal/instruments/isaac/tools/\\\hspace*{5mm}spectroscopic\_standards.html\#Telluric}. The
position of the telluric lines was fit with a $\mathrm{2^{nd}}$ order
polynomial which was used to provide the wavelength calibration. The
mean rms of the solutions was found to be approximately
4~$\mathrm{km\,s^{-1}}$, which was deemed acceptable. Only the spectra
of HD~327083 were used to provide wavelength calibration as the
spectra of the standards exhibit CO absorption features, which made
identification of the telluric lines less secure.

\smallskip

On several occasions, the closure phases of the calibrator objects
were non-zero. In an attempt to correct this, we subtracted the mean
of the calibrator closure phases from those of HD~327083. Since we
merged the OI files prior to frame selection and averaging, the
closure phase data calibration could be compromised by temporal
variations in the instrumental closure phase. In general, the closure
phases of the two calibrators are consistent, indicating this is not
an issue. However, the calibrator closure phases for the observations
conducted in September were inconsistent by approximately
5${\degr}$. This translates to a systematic uncertainty in the final
calibrated measurement.

\smallskip

The two calibrator stars, HD~151078 and HD~161068, were also used as
telluric standards. For each night of observing, we attempted to
remove the telluric absorption features in the spectrum of HD
327083. Before this could be done, the CO absorption features in the
spectra of the standard had to be removed \citep[see
e.g. W12,][]{Tatulli2008}. This was done by dividing the standard
spectra with a telluric corrected spectrum of a template star with the
same spectral type. High-resolution (R$\sim$18,000), $K-$band template
spectra were obtained from the library of 25 standard stars observed
with GNIRS provided by GEMINI
observatory\footnote{http://www.gemini.edu/sciops/instruments/nearir-resources/\\\hspace*{5mm}spectral-templates/r18000/}. For
HD~161058 (K4/K5III\footnote{from SIMBAD:
  http://simbad.u-strasbg.fr/simbad/}) the spectrum of HD~78541, also
of spectral type K4/5III was used. In the case of HD~151078 (K0III)
the average spectrum of HD~74137 and HD~28 (both K0III) was used.

\section{Observational results}

\label{res}

The AMBER and CRIRES observations spectrally resolve the CO bandhead
emission of HD~327083 for the first time (the spectra are presented in
Sects \ref{CO_var} and \ref{CO_mod_fit} respectively). The high
spectral resolution of CRIRES ($\sim$6~$\rm{km\,s^{-1}}$) and the high
SNR of the data ($\sim$200) provide a unique view of the bandhead
spectral profile. The bandhead clearly displays a blue wing, the
so-called ``blue shoulder'', which can indicate that the emission
originates in a rotating medium such as a disc \citep[see e.g.][
hereafter W10, and references therein]{MeCO}. No clear signature is
seen over the bandhead emission in the differential
visibilities. However, we report the first detection of a signal
across CO bandhead emission in AMBER differential phase observations
(this is presented in Sect. \ref{dp}). This allows us to probe the
emitting region with a precision as high as 30~$\mu$mas.

\smallskip

Turning to the closure phase measurements (presented in
Sect. \ref{cp}), no clear signature is seen over the CO bandhead
emission. In general, the average closure phases are close to zero;
which suggests that the environment of HD~327083 is relatively
symmetric. However, the observations in July suggest a shift to
non-zero values, which indicates an increase in the asymmetry of
the environment of HD~327083. In the following sections we analyse
these points in more detail to constrain the source of the CO emission
and the nature of the mass loss of HD~327083.

\section{Analysis and modelling}

\label{ana}

\subsection{CO bandhead spectra: assessment of variability}

\label{CO_var}

Our AMBER observations of HD~327083 span a period of approximately 5
months, which is more than 80 per cent of the orbital period estimated
by M2003. Therefore, we can use the observed flux spectra to assess
whether the CO emission is variable, which might indicate mass
transfer at particular orbital phases. To determine whether there are
significant differences between the average spectra from each night of
observation, we use the temporal variance spectrum \citep[TVS,
see][]{Fullerton1990}. The TVS can be expressed by:

\begin{equation}
(TVS)_{\lambda} \approx \frac{1}{N-1} \sum_{i=1}^{N}\Big[\frac{F_i(\lambda) - F_{\rm{av}}(\lambda)}{\sigma_i(\lambda)}\Big]^2	
\end{equation}

where N represents the number of spectra, $F_i(\lambda)$ denotes the
individual spectra, $F_{\rm{av}}(\lambda)$ is the average spectrum and
$\sigma_i$ is the uncertainty at each point of the individual
spectra. Here, we used the rms in the continuum as a constant
uncertainty across the individual spectra.

\smallskip

\begin{center}
  \begin{figure}
    \begin{center}
      \includegraphics[width=0.425\textwidth]{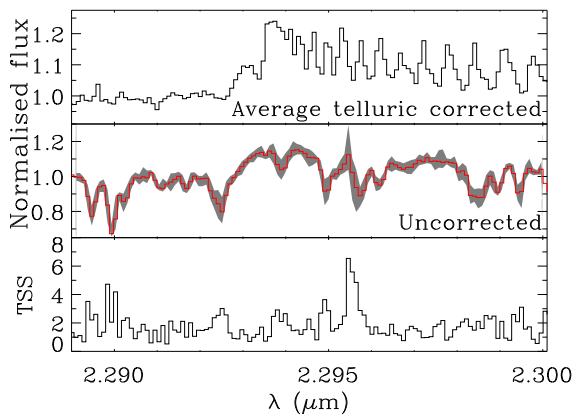}
      \caption{The upper panel presents the time-averaged, telluric
        corrected AMBER spectrum. The central panel contains the
        un-processed (not corrected for telluric absorption) AMBER
        spectrum. The shaded region represents the scatter of the six
        average spectra constructed for each observing date. The lower
        panel presents the temporal sigma
        spectrum. \label{FIG:spec_tvs}}
    \end{center}
  \end{figure}
\end{center}

Following the example of \citet{Oudmaijer1994}, we calculated the
temporal sigma spectrum, TSS, which is given by $TSS=\sqrt{TVS}$. This
value is expressed in terms of the noise level. Therefore, if there
are no significant deviations between the individual spectra, this
value will be approximately unity. On the other hand, significant
variations will be associated with a large TSS value. The average
AMBER spectrum of HD~327083 (uncorrected for telluric absorption) and
the associated TSS is shown in Fig. \ref{FIG:spec_tvs}. HD~327083 was
observed at different air-masses each of the six times it was
observed. As a result, the strength of the telluric absorption
features in each spectra is different, although, as shown in
Fig. \ref{FIG:spec_tvs}, the differences are relatively small. The
telluric lines in the average spectrum are associated with peaks in
the TSS. This confirms that this approach is sensitive to
variability. The bandhead itself, however, exhibits no such
feature. Therefore, we conclude that these data contain no evidence of
variability in the CO bandhead emission.

\smallskip

We note that the final telluric corrected AMBER and CRIRES spectra are
essentially consistent, which is to be expected if the emission is not
variable. However, comparison of the two suggests that the telluric
correction in the AMBER spectrum is not perfect. This is likely
because the standards were chosen as interferometric calibrators
rather than telluric standards and thus they do not necessarily
provide the best air-mass match. {{From here-on, as far as flux spectra
are concerned, we concentrate on the high spectral resolution CRIRES
spectrum.}}

\subsection{Modelling the CO flux spectrum}

\label{CO_mod_fit}

With the aim of constraining the origin and kinematics of the bandhead
emission, the observed spectral profile is fit with a kinematical
model. Specifically, to evaluate the disc scenario, we attempt to fit
the CO bandhead with a model of a rotating disc. We have already used
a similar model to fit the CO emission of massive young stellar
objects in other publications (W10, Ilee et al. in prep.), and we thus
only describe the methodology briefly here.

\smallskip

A simple, geometrically flat disc in Keplerian rotation is
constructed. The excitation temperature and surface number density
decrease with increasing radius according to power laws. The
temperature decreases with $R^{-0.75}$ and the surface number density
falls off with $R^{-1.5}$, in line with standard disc theory
\citep[see e.g.][]{Kraus2000}. We note that the temperature is the
excitation temperature used to determine the relative populations of
the CO ro-vibrational energy levels. Therefore, it may not necessarily
follow the thermal temperature distribution. For example, it is
possible the vibrational levels are not in LTE
\citep{Najita1996,Martin1997}. The excitation temperature distribution
is difficult to constrain with only one bandhead profile as it can
vary significantly \citep[see e.g.][]{Berthoud2007}. Therefore, we fix
the temperature distribution as described.

\smallskip

The temperature and surface density at the inner radius,
$R_{\rm{in}}$, are free parameters. The CO emission is assumed to
originate in the region of the disc where $1000~K \le T \le
5000~K$. In the case of HD~327083, we set the inclination of the disc to
be $48.5\degr$, as determined by W12. In most cases, it would be
assumed that the disc extends from the stellar surface. However, in
the case of HD~327083, it is possible that the disc is circumbinary
and the inner disc is cleared. Therefore, we allow the inner rim of
the disc to vary. The free parameters of this particular model are
thus: the inner rim of the disc (technically the radial distance of
the smallest annulus), the temperature at $R_{\rm{in}}\rm $, the intrinsic
line-width and the surface number density at $R_{\rm{in}}\rm $. The disc was
split into radial and azimuthal cells and the CO emission of each cell
was calculated according to the methodology of
\citet{Kraus2000}. Finally, once the individual cell spectra were
calculated, they were summed to create the total spectrum. This was
then smoothed to match the spectral resolution of the observations.

\smallskip

This relatively simple approach is standard in such studies as there
are no sophisticated models including CO bandhead emission
available. As the code is analytic, it is readily used in a fitting
scheme. The best-fitting model was determined using the downhill
simplex algorithm (implemented in {\sc{idl}} as the amoeba
routine). Once the best fit model was found, monochromatic images were
created, which were subsequently compared to the interferometric data.
The uncertainties associated with the final parameters were determined
in the following way. A 2D reduced $\chi^2$ map was constructed for
each set of parameters, centred on the best fitting model. The
uncertainties in the best fitting parameters were estimated by
determining the maximum parameter change that results in a change in
$\chi^2$ of 1. As discussed in \citet{DS2011}, this likely results in
an over-estimation of the uncertainty (provided the data points are
un-correlated). To investigate the possibility of additional fits with
different inner radii, we constructed 2D $\chi^2$ maps centred on the
best fitting model but displaced in the direction of the inner radius
by $-5$ and $+5~R_{\star}$ (we concentrated on the inner radius as
this influences the interpretation of the data, as discussed in
Sect. \ref{disc}). In both cases, the minimum $\chi^2$ was found to be
greater than that of the model presented in Table
\ref{Table:CO_fit}. Therefore, we conclude that the best fitting model
does indeed correspond to the global $\chi^2$ minimum.

\smallskip

The model spectrum that is found to best reproduce the CO bandhead
profile is shown in the top panel of Fig. \ref{FIG:CO_fit}. The
parameters of the best fitting model and the adopted stellar
parameters are presented in Table \ref{Table:CO_fit}. In general, the
model provides an excellent match to the features of the
bandhead. Indeed, the fit is remarkably good considering the limited
number of free parameters and the simplicity of the model. Therefore,
we conclude that a model of a Keplerian disc with an inner radius of
approximately 3~AU and a relatively low initial temperature,
$T_0$$\sim$1700~K, appears to be a fitting representation of the CO
emitting region associated with HD 327083.

\smallskip

However, the possibility remains that the source of the CO emission is
not rotationally dominated. Indeed, the favoured model of the
circumstellar environment of sgB[e] stars features an equatorial
outflow. To investigate whether this scenario is applicable to
HD~327083, we perform an additional fit to the CRIRES spectrum with an
outflow model.

\smallskip

The standard model for the sgB[e] environment is a dense equatorial
outflow and a faster wind concentrated in the polar regions. The
equatorial outflow is thought to be the result of rapid rotation and
the bi-stability jump leading to enhanced line driven mass loss in the
equatorial region \citep[see e.g.][]{Pelupessy2000}. In modelling the
emission from an equatorial outflow, we adopt the analytical treatment
of \citet{P2003} who compared the SEDs of disc and outflow models to
observations of the sgB[e] star R126. Specifically, the CO emission is
modelled as originating in a flat structure containing both rotation
and outflow. The velocities are given by:
\begin{equation}v_{\rm{Rot}}=\sqrt{GM_{\star}/R_{\star}}\left(\frac{r}{R_{\star}}\right)^{-1}\end{equation}
and
\begin{equation}v_{\rm{Out}}=v_{\rm{Sound}}+(v_{\infty}-v_{\rm{Sound}})\left(1-\frac{R_{\star}}{r}\right)^{\beta}\end{equation}
which represent angular momentum conserving rotation and the velocity
structure in a line driven wind respectively. 

\smallskip

We fit the CRIRES spectrum with a geometrically flat, equatorial
outflow model featuring both these velocity laws by treating
the terminal velocity, line-width, initial number density and
temperature as free parameters and setting $R_{\rm{in}}=1~R_{\star}$
(since the wind is launched from the star) and $\beta=0.8$ (which is
appropriate for a hot star wind). The inclination was also set, as in
the disc modelling. The temperature and number density distributions
follow the same power laws as in the disc model, which is appropriate
for this analytic treatment \citep[see e.g. the discussion on the
heating of circumstellar discs and winds in][]{P2003}. As before, the
best fitting model was found using the downhill simplex algorithm.

\smallskip

The best fitting spectral profile of the equatorial outflow model is
also shown in the upper panel of Fig. \ref{FIG:CO_fit}. The best
fitting parameters are: $N_{\rm{0}}=2.7\times10^{22}{\rm{cm^{-2}}}$,
$T_{\rm{0}}=19624$~K, $Linewidth=6.8$~$\rm{km\,s^{-1}}$ and
$V_{\infty}=69.4$~$\rm{km\,s^{-1}}$. As can be seen, the outflow model
can approximately reproduce the CRIRES spectrum, although the quality
of the fit is noticeably worse than that provided by the disc
model. This is reflected in the large reduced $\chi^2$, which is
$\sim$100. Therefore, the $\chi^2$ values favour the Keplerian
model. Nonetheless, since the spectrum is approximately recreated, it
is not possible to confidently discard the outflow model based on the
CO bandhead profile alone. However, the addition of our AMBER data
provides another opportunity to differentiate between the two
models. This is described in the following section.

\begin{center}
  \begin{table}
    \begin{center}
      \caption{{{The parameters of the CO emitting Keplerian disc model that best fits the CRIRES spectrum.}}\label{Table:CO_fit}}
      \begin{tabular}{l c c}
        \hline
        Parameter & Value & Notes \\
        \hline
        \hline
        Cent. Mass  &  25~$M_{\odot}$  & Set$^1$ \\
        $R_{\star}$  &  25~$R_{\odot}$  & Set$^2$ \\
        $i$  &  45.8  & Set$^3$ \\

        $N_{\rm{0}}$ & $0.4\pm1.8\times 10^{24}$~$\mathrm{cm^{-2}}$& \\
        $R_{\rm{in}}$  & 25.5$\pm$2.3~$R_{\star}$ & $\sim$3~AU, 2~mas\\
        $Linewidth$   & 3.2$\pm$3.2~$\mathrm{km\,s^{-1}}$& \\
        $T_{\rm{0}}$ & 1721$\pm$540~K& \\

        $\chi^2$ & 3.96 & \\
        
        \hline

        \end{tabular}
        \tablefoot{1: Based on an early B type primary and an F type companion (M2003), 2: This sets the minimum possible inner disc radius, based on the values reported by M2003, 3: Set by our previous AMBER observations (W12) and assuming that the CO emission is co-planar with the continuum emission. The outer radius is simply the radius at which the temperature falls below 1000~K, and in this case is approximately 6~AU.}
      \end{center}
  \end{table}
\end{center}

\subsection{Differential visibilities and phases}

\label{dp}

The spectral modelling of the previous section provides detailed
predictions regarding the spatial distribution of the CO
emission. Therefore, the AMBER data offer a unique test of the
spectral models that fit the CO bandhead. No clear signature is seen in
the differential visibilities across the bandhead
emission. Considering the uncertainties in the visibilities (typically
10 per cent) and the low line-to-continuum flux ratio ($\sim$0.2),
this is consistent with the CO bandhead emission originating close to,
and perhaps interior to, the continuum emission (see the discussion in
W12). Our spectral models place the CO material interior to, but
extending to, the continuum emitting region. Therefore, the best
fitting models are consistent with the observed visibilities

\smallskip

A second, more stringent test is provided by the AMBER differential phase
measurements. Differential phases are sensitive to photo-centre
displacements associated with asymmetric sources of flux, such as
rotating discs \citep[see e.g.][]{Lachaume2003}. Therefore, we compare
the differential phase measurements to the predictions of the models
that fit the CRIRES data. To calculate the photo-centre offsets of
the models, it is assumed that the continuum is centro-symmetric. Most
of the closure phases measured are close to zero (see Sect. \ref{cp}),
indicating that the environment of HD~327083 is relatively symmetric,
justifying this approach. 

\smallskip

To improve the SNR of the data, we re-bin the differential phases by a
factor of 3 (approximately a resolution element). The minimum
uncertainty was $\sim$30~$\mu$as. This uncertainty corresponds to the
data obtained with the 34.7~m D0-A1 baseline used on $\mathrm{7^{th}}$
May 2011. These data alone are precise enough to test the models at a
significant level. The errors on the other baselines are generally a
factor of two or more greater, so we concentrate on the data obtained
with this baseline only.

\smallskip

\begin{center}
  \begin{figure*}
    \begin{center}
      \begin{tabular}{l l}
      \includegraphics[width=0.425\textwidth]{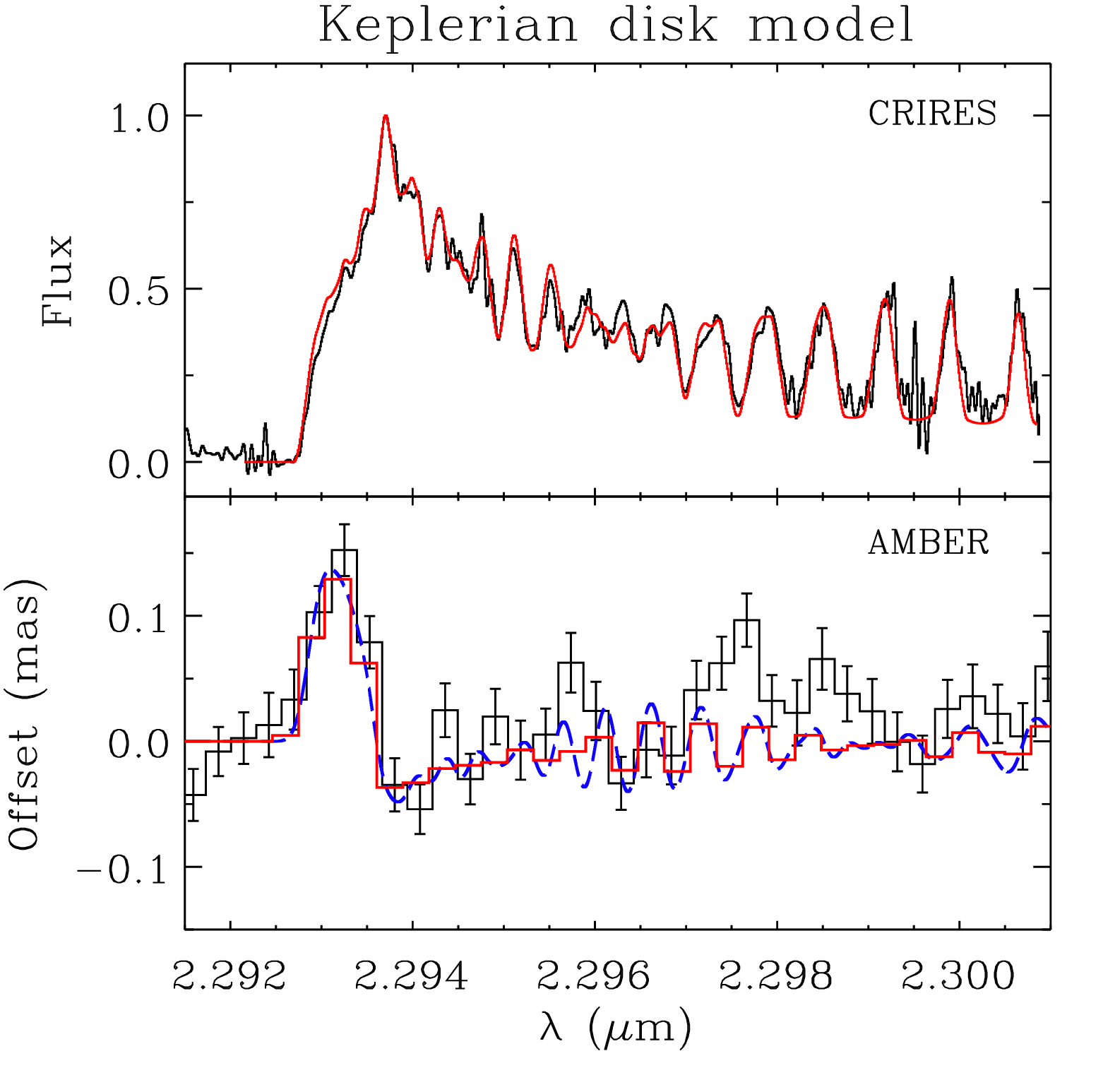} & 
      \includegraphics[width=0.425\textwidth]{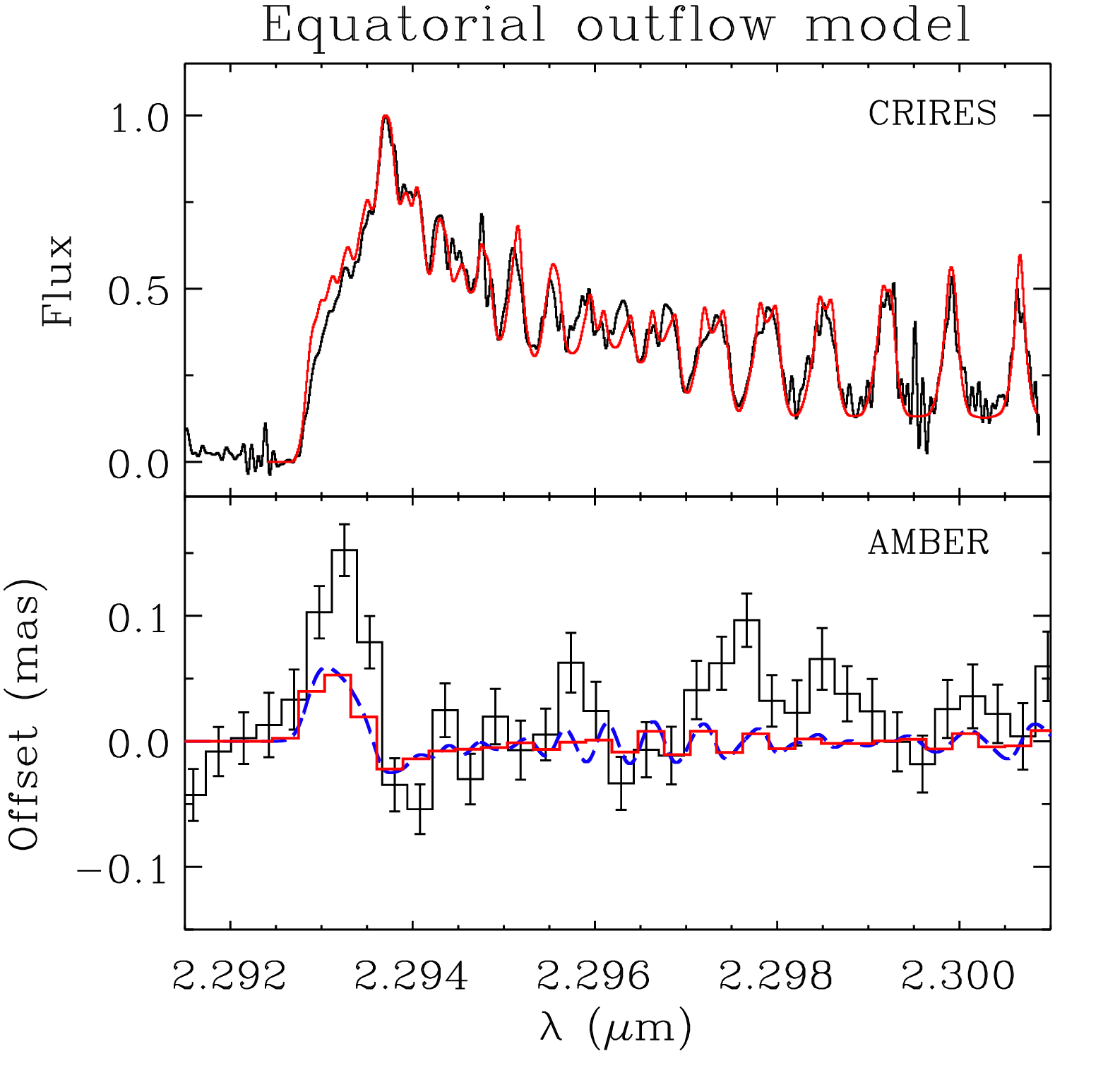}\\
    \end{tabular}
    \caption{A comparison between the observations and the two
      physically different best fitting models explored in the
      text. The upper panels present the average CRIRES spectrum and
      the profile of the best fitting model. In the lower panels, the
      observed photo-centre offsets calculated using AMBER
      differential phase measurements from the baseline with the
      lowest noise level (D0-A1 used on the $\mathrm{7^{th}}$ of May)
      are shown. For each model that fits the CRIRES spectrum we
      over-plot the predicted signature for the selected baseline. The
      model signatures are smoothed to the spectral resolution of
      AMBER and are shown re-binned using the size of the AMBER pixels
      and with a finer sampling.\label{FIG:CO_fit}}
    \end{center}
  \end{figure*}
\end{center}

\begin{center}
  \begin{figure*}
    \begin{center}
      \begin{tabular}{l l l}
        \includegraphics[width=0.3\textwidth]{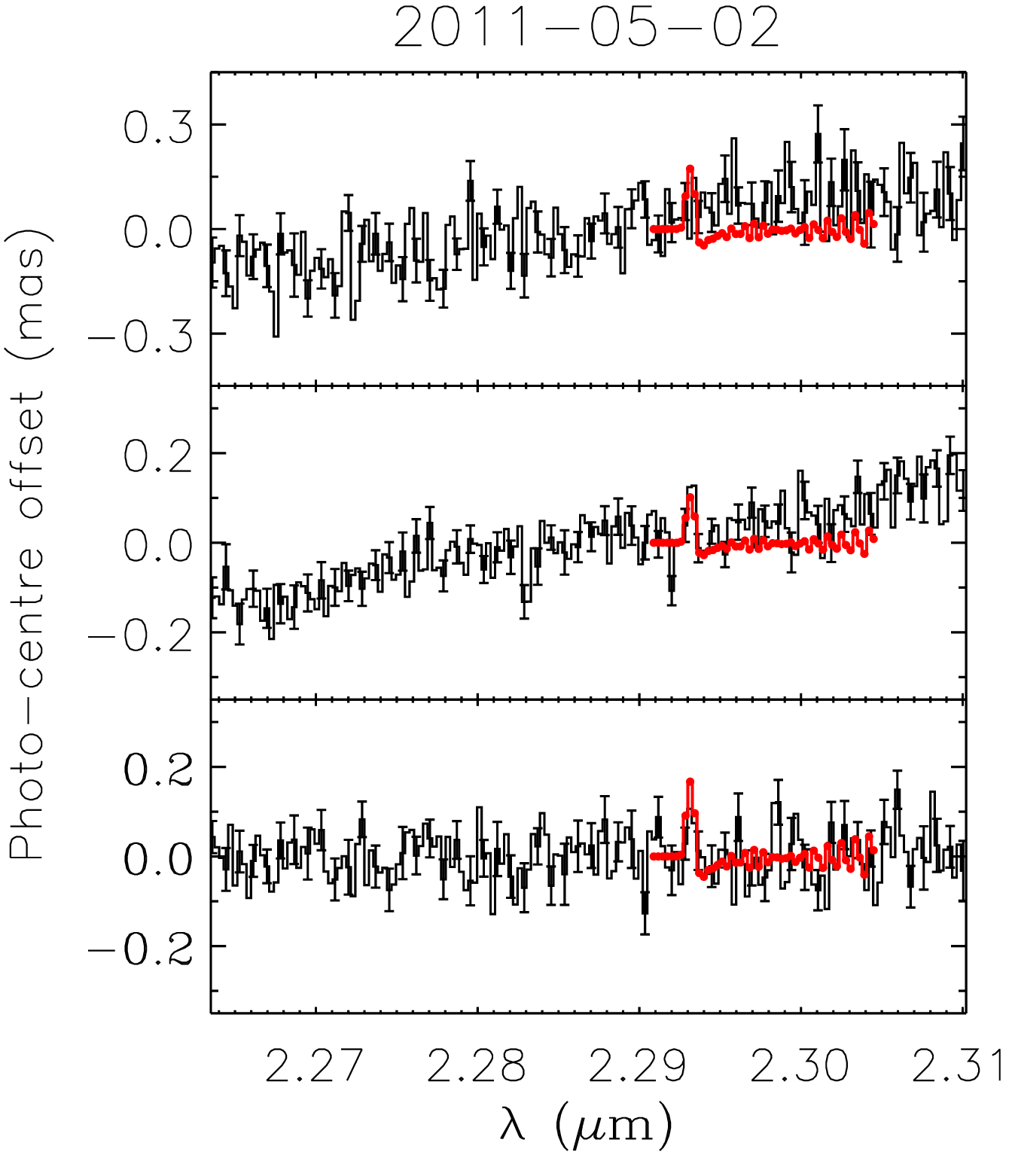} & 
        \includegraphics[width=0.3\textwidth]{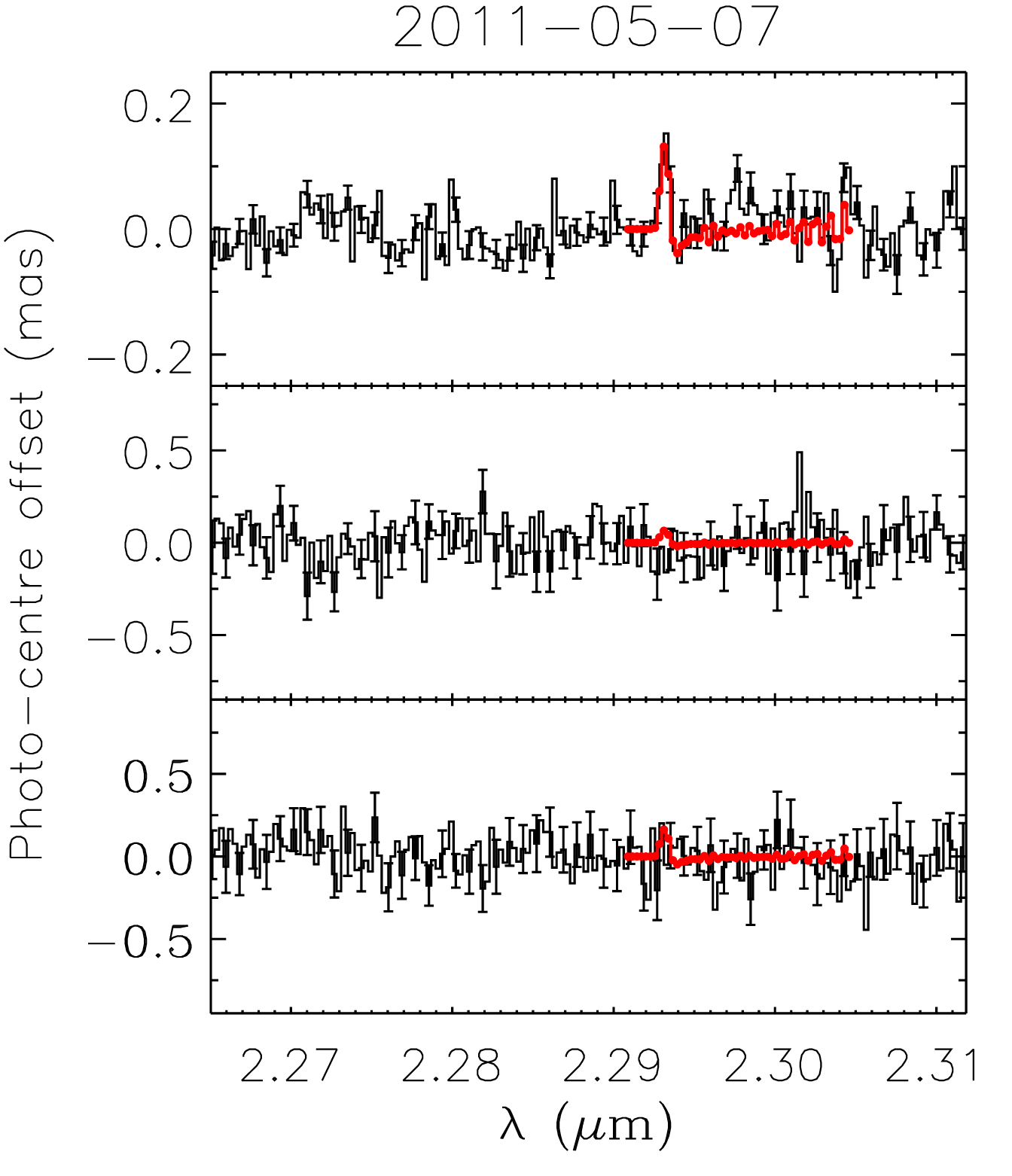} & 
        \includegraphics[width=0.3\textwidth]{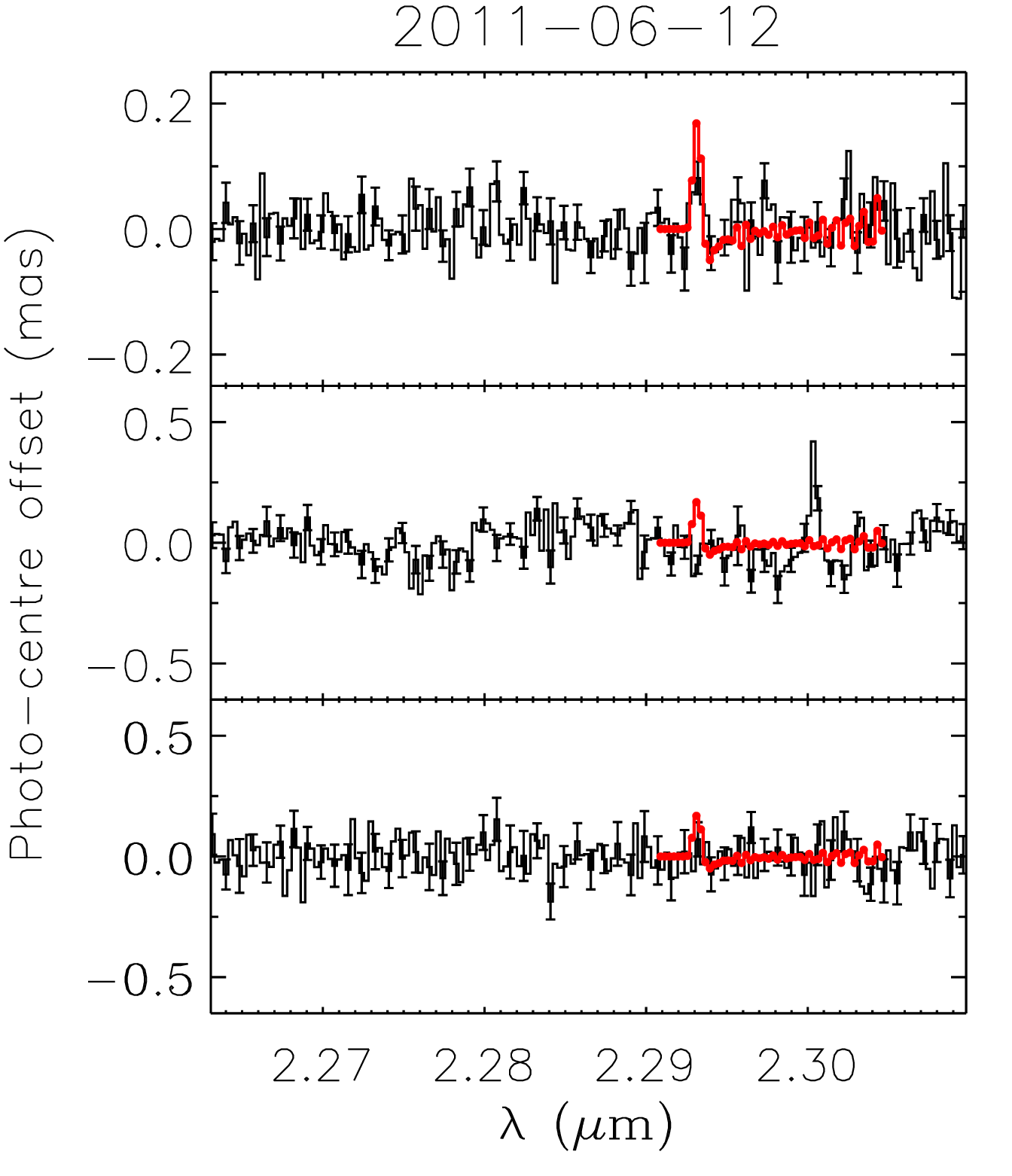} \\

        \includegraphics[width=0.3\textwidth]{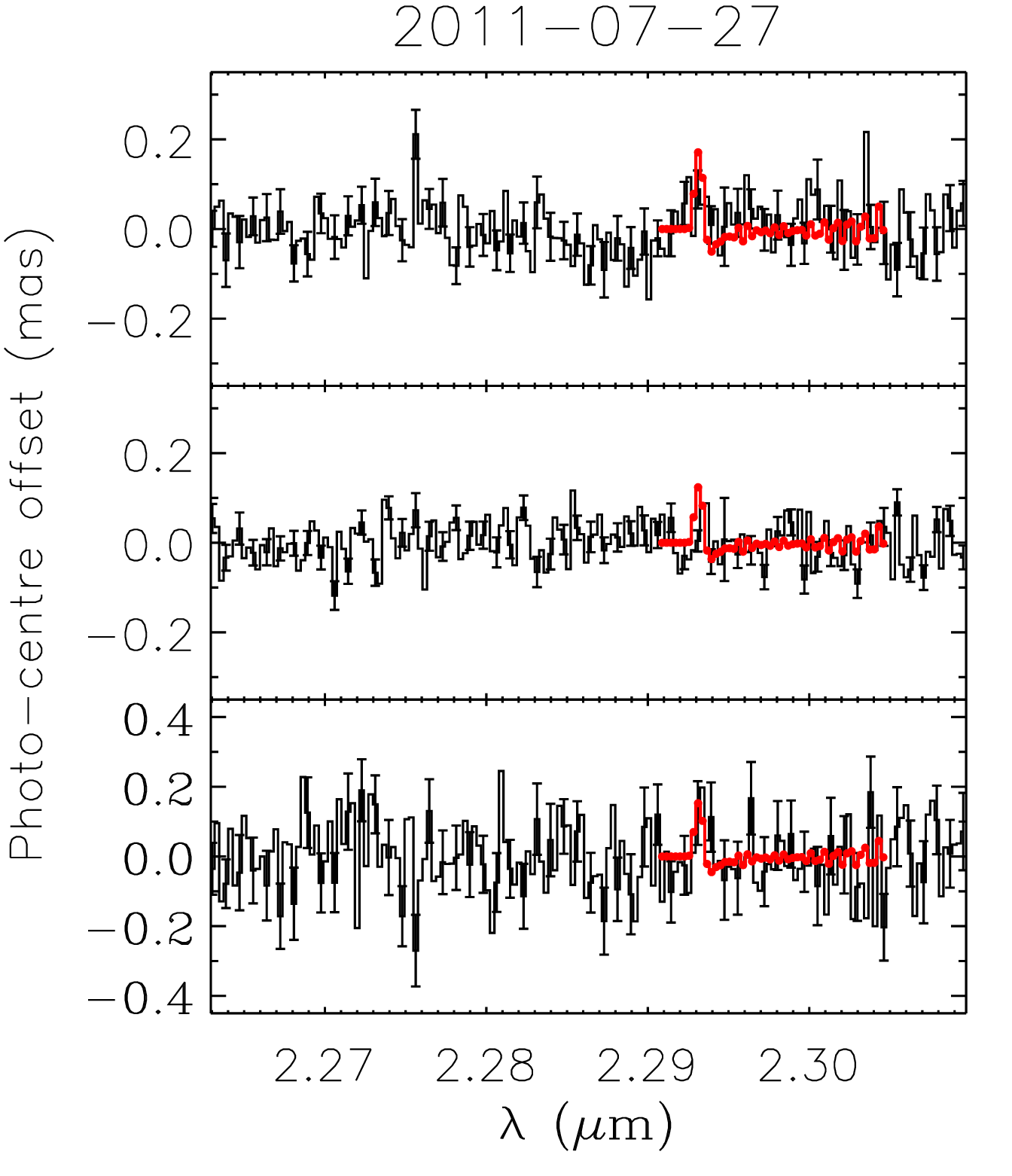} & 
        \includegraphics[width=0.3\textwidth]{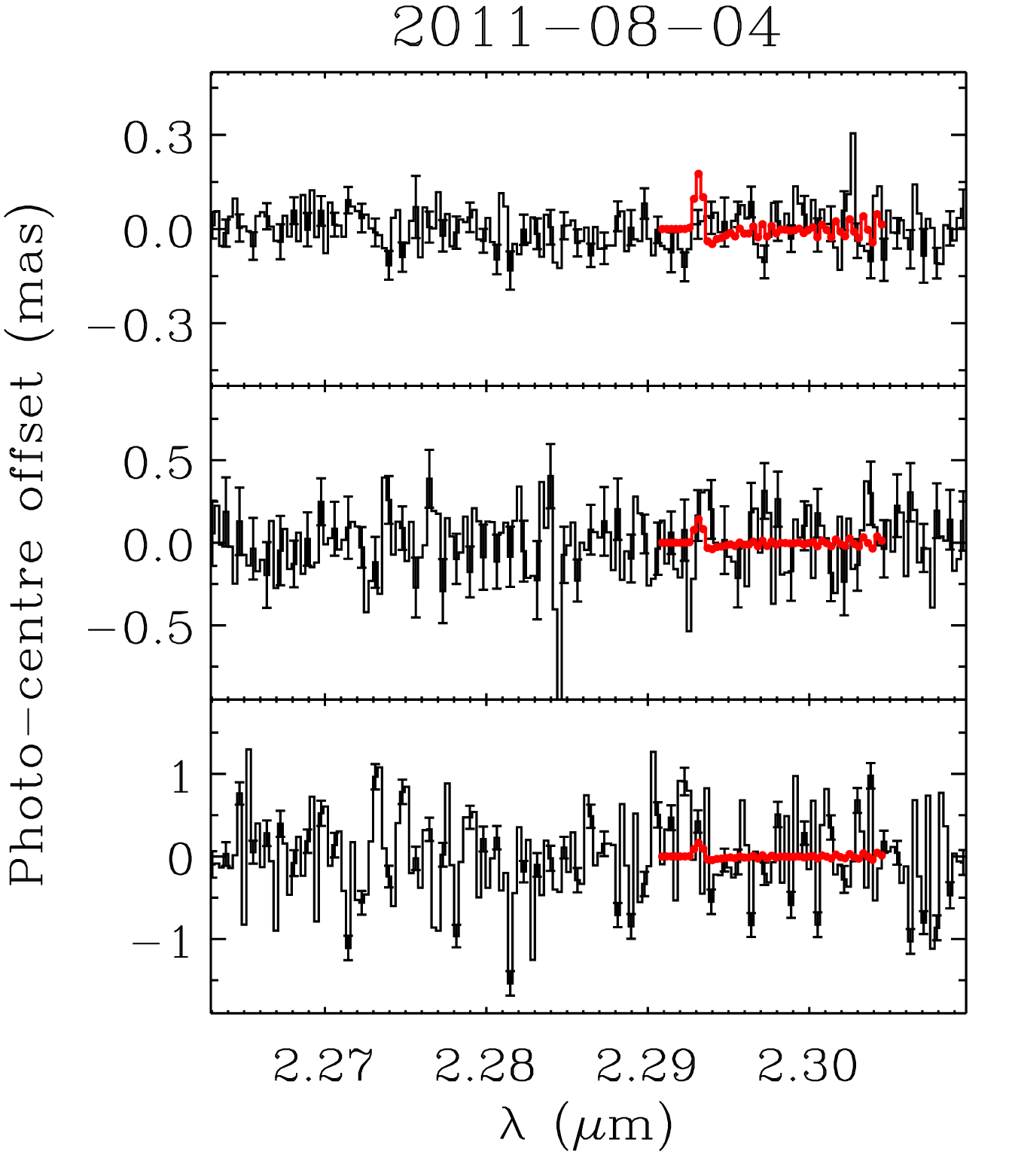} & 
        \includegraphics[width=0.3\textwidth]{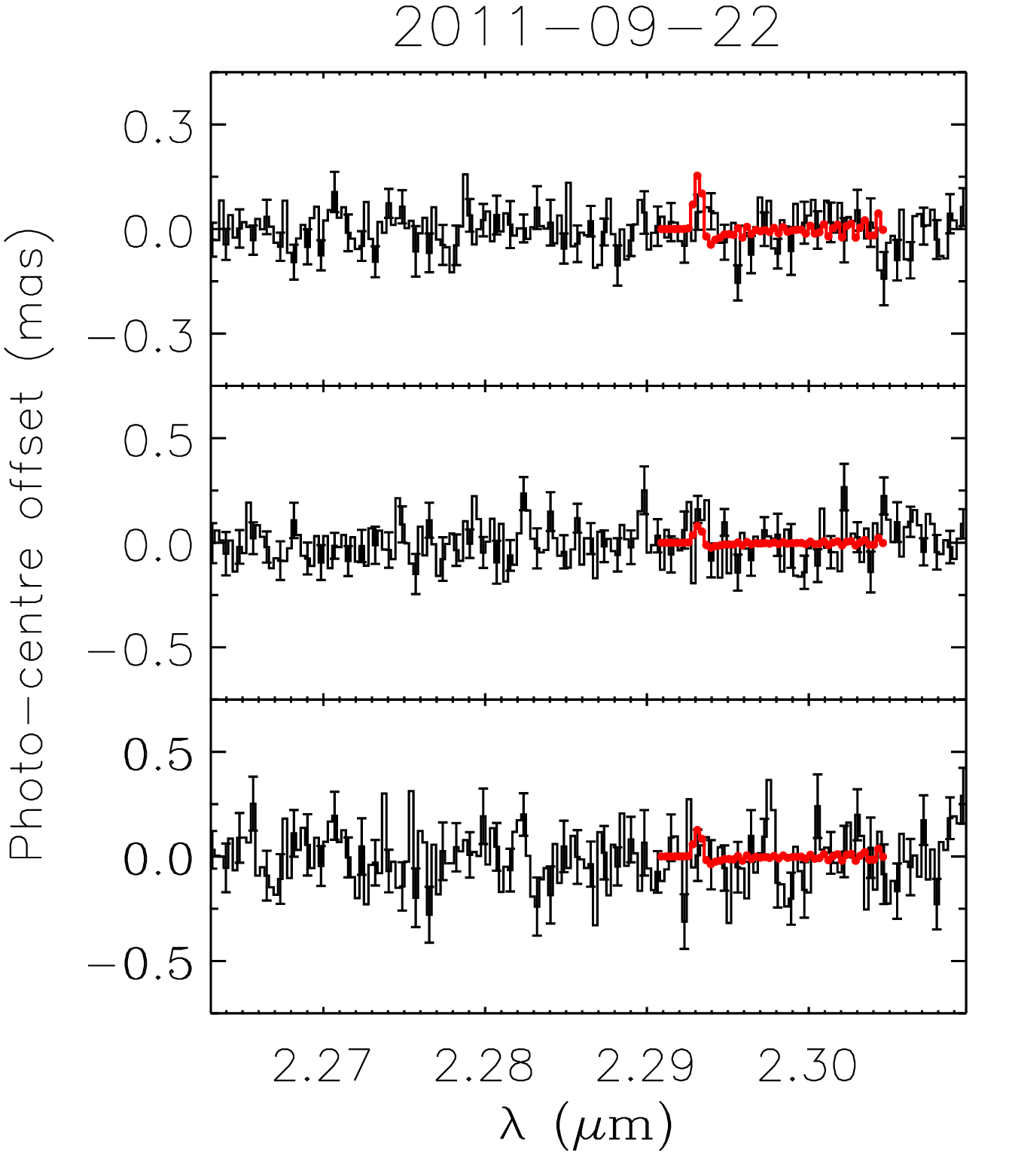} \\

      \end{tabular}
      \caption{Average differential phases converted into photo-centre shifts for each night of observing. The predictions of the disc model that fits the CRIRES spectrum are over-plotted.\label{FIG:diff_phase_1}}
    \end{center}
  \end{figure*}
\end{center}

A comparison between the predicted photo-centre offsets and that
calculated from the differential phase measurements using the
aforementioned baseline is presented in the lower panel of
Fig. \ref{FIG:CO_fit}. The offsets calculated from the differential
phases across the other baselines are presented in
Fig. \ref{FIG:diff_phase_1}. The combination of high spatial and
spectral precision provides stringent constraints on the kinematics of
emission line regions. As can be seen, the predicted offsets of the
best fitting disc and outflow models are different and thus we can use
the spatial information provided by the differential phases to
differentiate between the competing models.

\smallskip

We focus on the best fitting disc model first. As can be seen in
Fig. \ref{FIG:CO_fit}, the maximum predicted offset occurs over the
bandhead shoulder where many individual ro-vibrational lines are
super-imposed (as discussed in W10). The offsets over the individual
ro-vibrational lines are clearly too small ($<$0.1~mas) to be detected
here. However, the predicted offset over the bandhead shoulder,
$\sim$0.15~mas, is large enough to detect with the current data, which
has an noise level of $\sim$30~$\mu$as. The observations display a
displacement of approximately 0.15~mas over the bandhead shoulder, in
excellent agreement with the prediction of the model that fits the
CRIRES spectrum. The significance of the peak pixel of the observed
signature is approximately 5~$\sigma$.  The agreement between the
amplitude of the model and observed signatures was optimised by
slightly changing the position angle of the disc (by 10\degr$\,$ so
that the difference between the position angle of the major axis of
the disc and the baseline was $\sim$42$\mathrm{^{\circ}}$). This
slight difference in PA could represent uncertainty in the previous
estimate. However, it should be noted that the distance to HD 327083
has an uncertainty of approximately 30~per cent, which translates to a
similar uncertainty in the model photocentre shift. Therefore, the
agreement between the model and observations could also be optimised
by slightly reducing the distance to HD 327083. As a result, the minor
change in PA is not considered significant.

\smallskip

We emphasise that bar a modest adjustment of the PA, the observed
differential phase signature was not fit. The model signature was
predicted by the fit to the CRIRES spectrum. The excellent agreement
between the model and the data in both the spectral and spatial
domains strongly suggests the model is an accurate representation of
the CO emitting region of HD 327083. We note that if the profile were
fit with rotation that decreased more rapidly with radius than the
Keplerian case, for example angular momentum conserving rotation in
which $v \propto r^{-1}$, the resultant positional offset would be
smaller than that observed and {\emph{vice versa}} \citep[see
  e.g.][]{Me_beta_CMi}. Therefore, the agreement between the model and
data is compelling evidence that the CO bandhead emission traces a
Keplerian disc.

\smallskip

\begin{center}
  \begin{figure*}
    \begin{center}
      \begin{tabular}{l l}
      \includegraphics[width=0.35\textwidth]{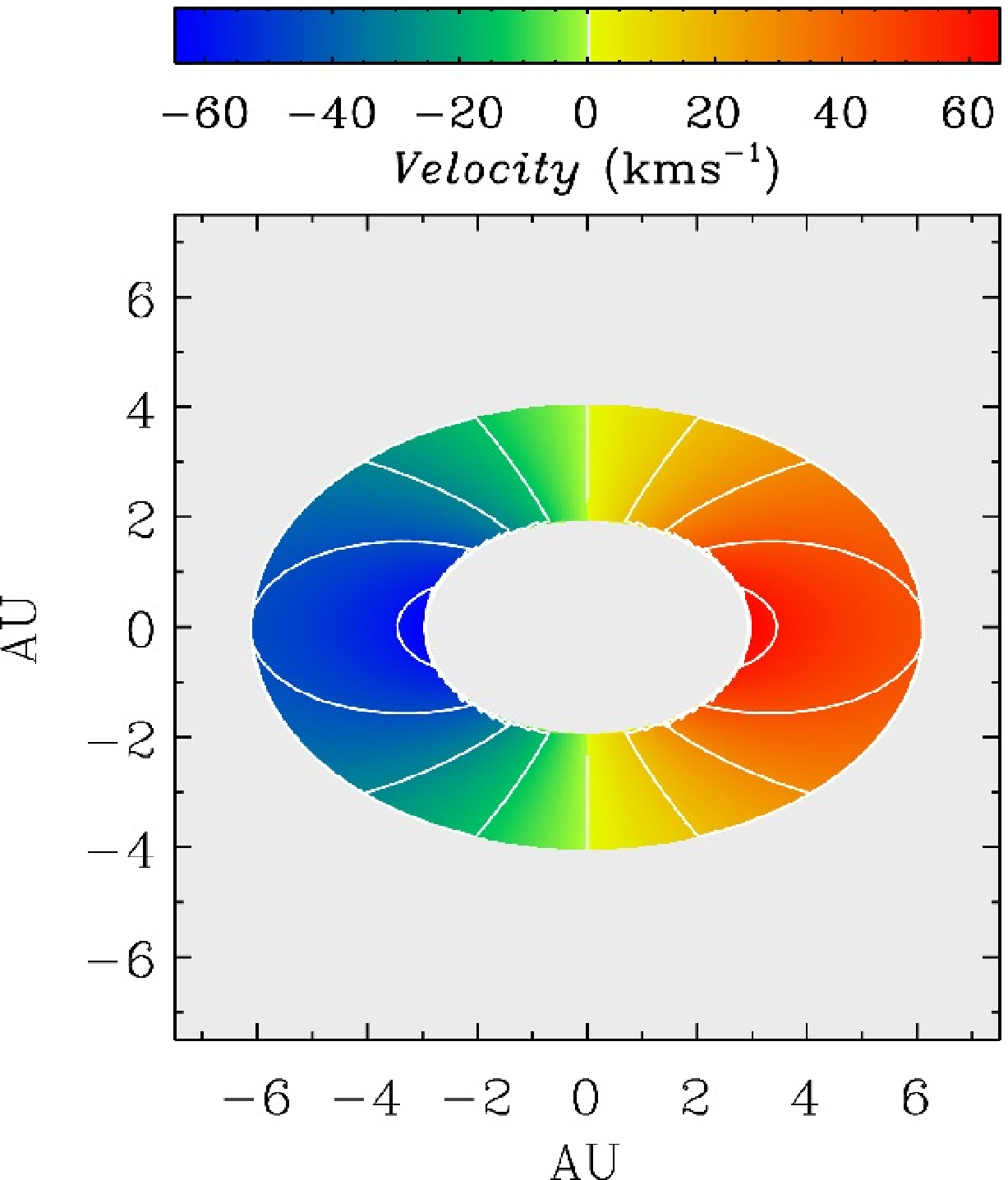} & 
      \includegraphics[width=0.35\textwidth]{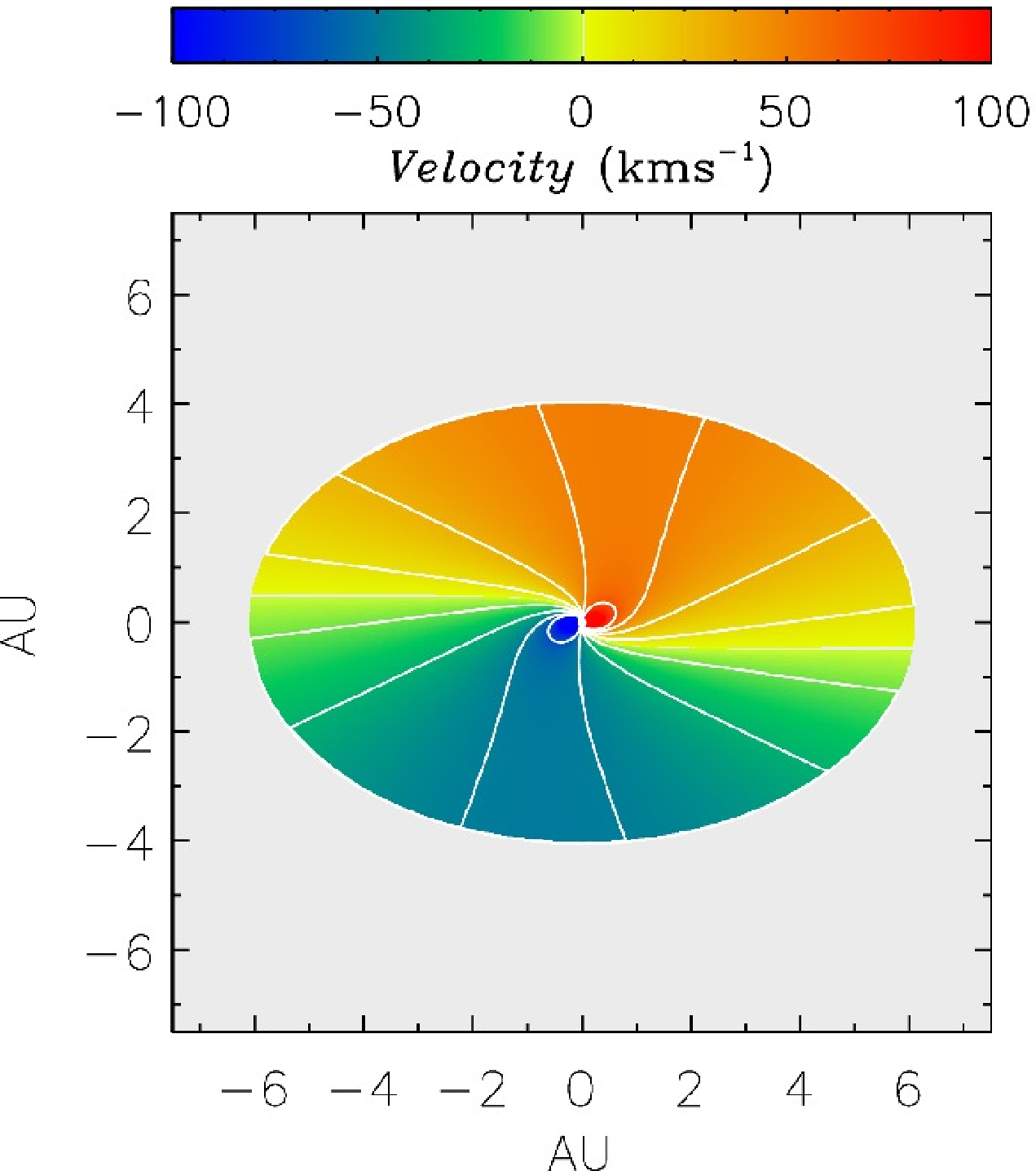} \\ 
      \includegraphics[width=0.345\textwidth]{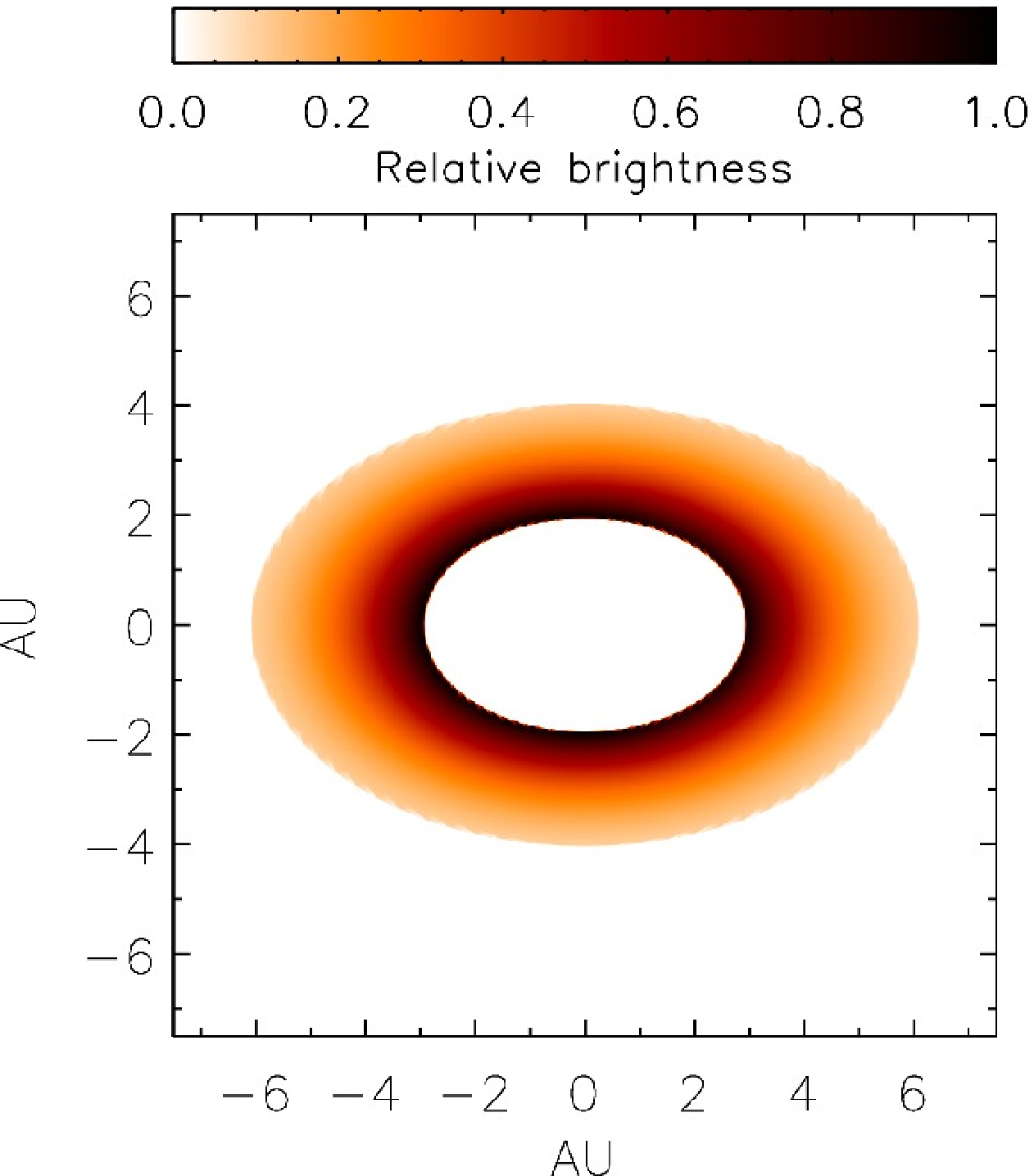} & 
      \includegraphics[width=0.345\textwidth]{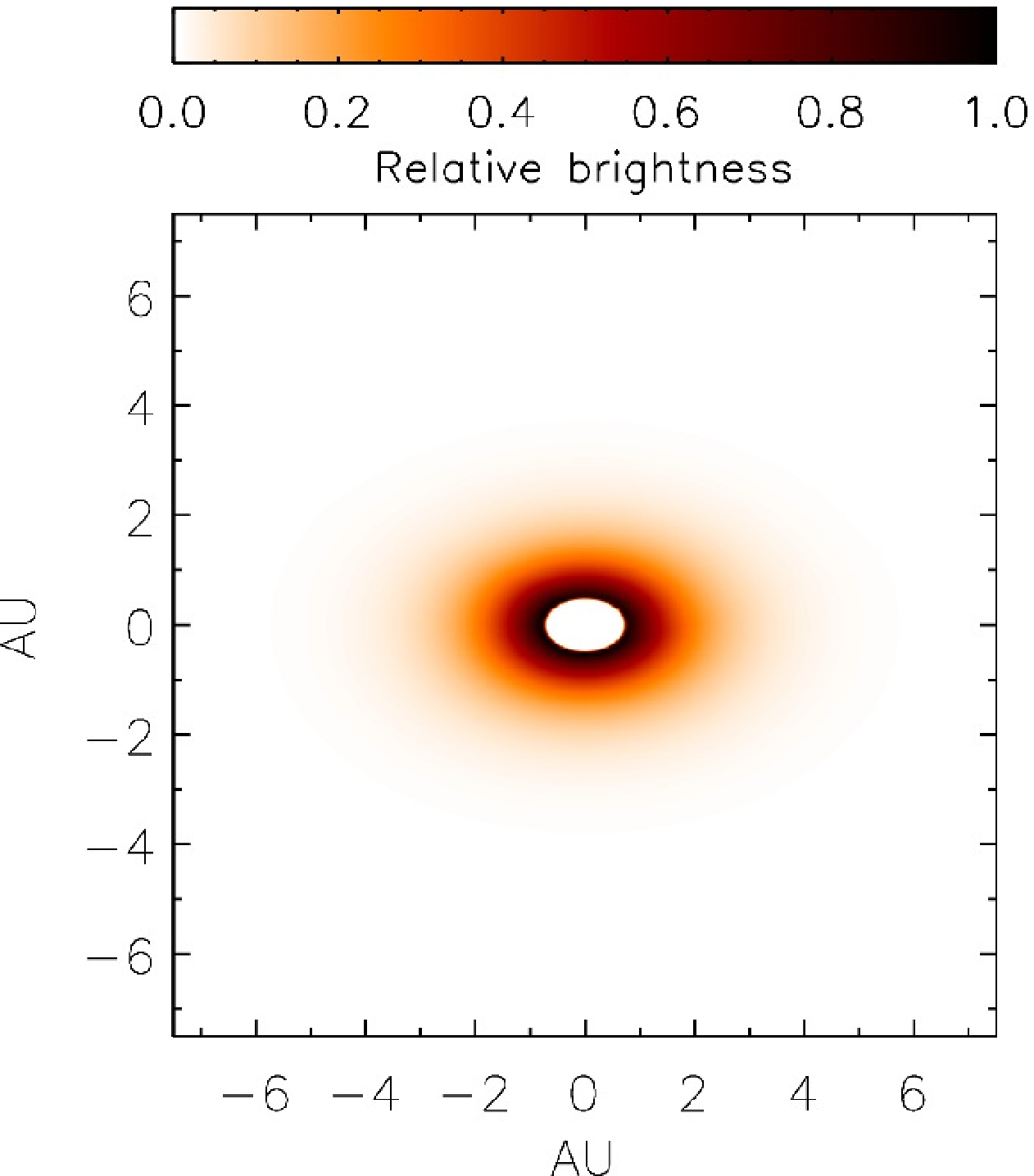} \\
      \end{tabular}
      \caption{A comparison between the velocity structure (upper row) and brightness distribution (lower row) of the best fitting disc (left column) and equatorial outflow (right column) models. The contours in disc velocity image are drawn at:  15, 30, 45 \& 60~$\rm{km\,s^{-1}}$ and the contour levels of the outflow velocity image are: 10, 30, 50 \& 70~$\rm{km\,s^{-1}}$. The maximum velocity shown in the outflow image is 100~$\rm{km\,s^{-1}}$, but the velocity at the inner edge of the structure is $\sim$400$\rm{km\,s^{-1}}$. The brightness distributions are normalised to the contribution of the brightest element.\label{FIG:model_comp}}
    \end{center}
  \end{figure*}
\end{center}

Unlike the disc model, the outflow model cannot reproduce the data
well. As shown, the photo-centre offset associated with the best
fitting outflow model is smaller than that of the best fitting disc
model. This is because the velocity structure and brightness
distribution of the two best fitting models are very different. This
is shown in Fig. \ref{FIG:model_comp}. The CO emitting region of the
outflow is clearly smaller than in the disc case. As a result, the
spatial signature of the outflow is smaller than that of the disc and
does not reproduce the differential phase shift seen (see
Fig. \ref{FIG:CO_fit}). The discrepancy between the model and
observations is too large to be accounted for by a modest change in PA
or distance. We conclude that the large $\chi^2$ of the spectral fit
and the poor reproduction of the photo-centre offset imply that the
source of the CO emission is not an equatorial outflow and that these
data favour the disc scenario.

\smallskip

\subsection{Closure phases}
\label{cp}

\begin{center}
  \begin{figure}
    \begin{center}
      \begin{tabular}{c}
      \includegraphics[width=0.425\textwidth]{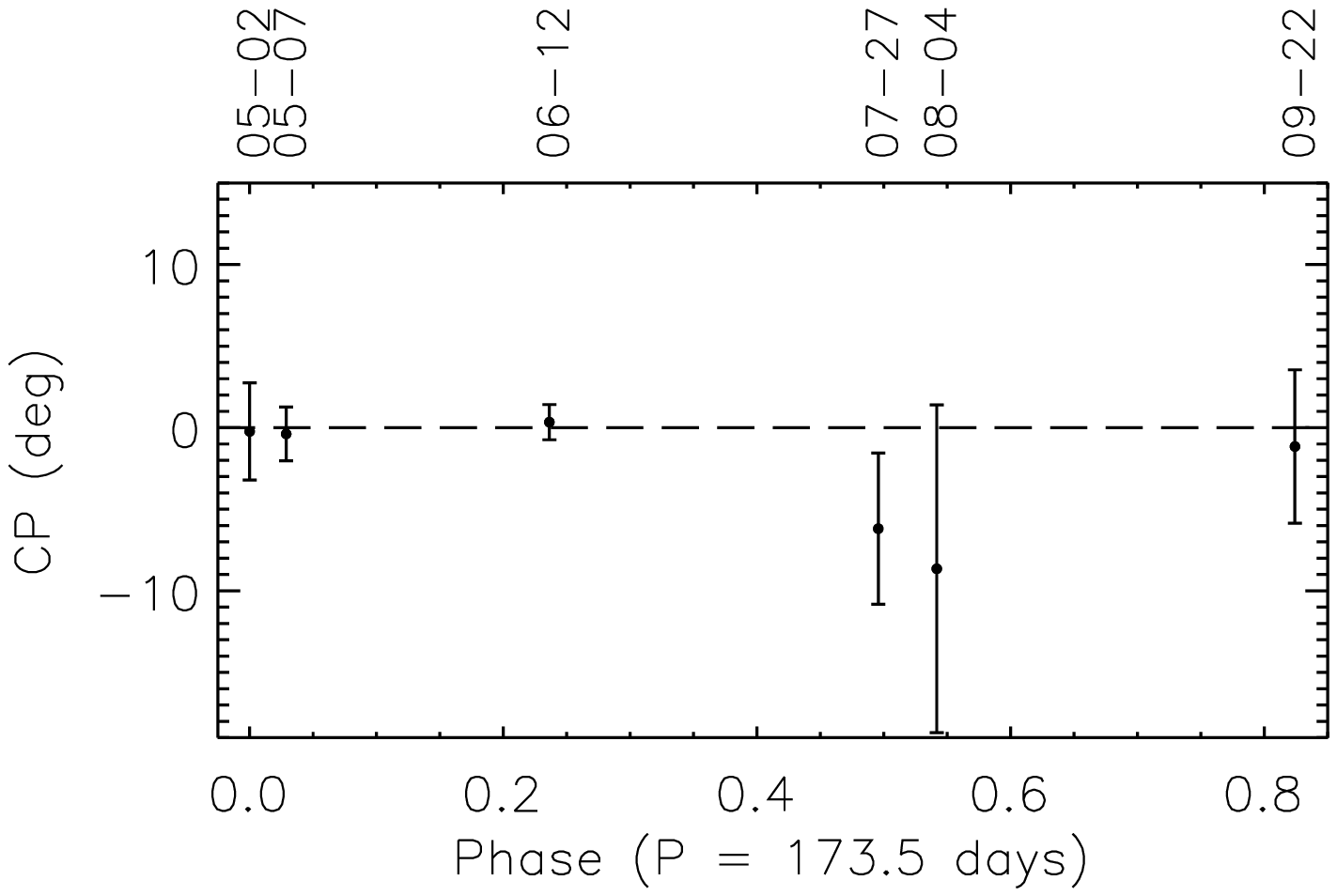} \\
      \includegraphics[width=0.425\textwidth]{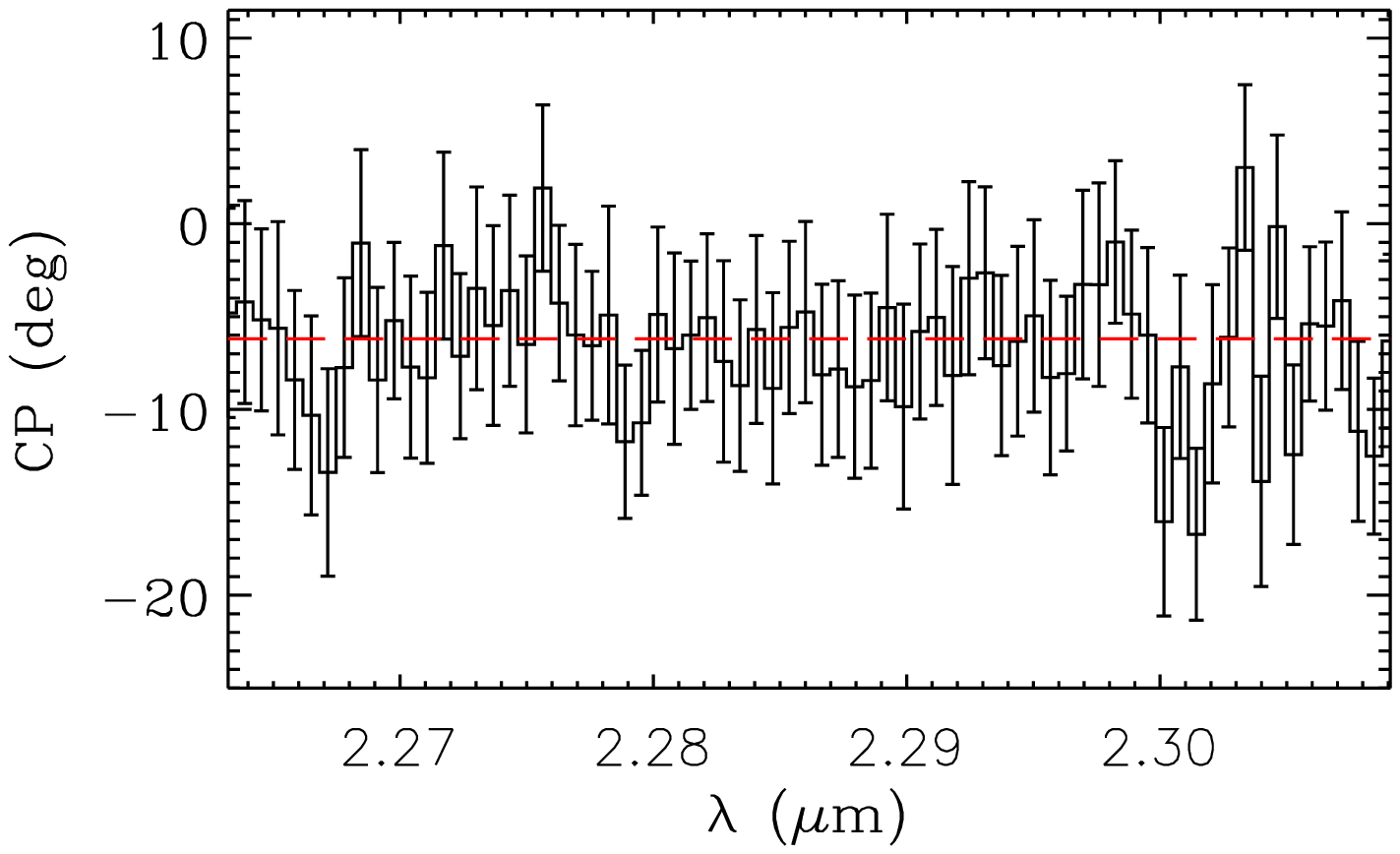} \\
      
      \end{tabular}
      \caption{The wavelength-averaged calibrated closure phase for
        each observing date (top panel) and an example of the closure phase data
        (data obtained on the $\mathrm{27^{th}}$ of June, lower panel). The data
        have been re-binned by approximately twice the resolution
        element.\label{FIG:VAR_CP}}
    \end{center}
  \end{figure}
\end{center}

The AMBER closure phase signal, sampled at different phases of the
binary period \citep[$\sim$6~months,][]{Miro2003}, has the potential to
trace changes in the degree of symmetry in the environment of HD
327083. This might be expected due to changes in the binary separation
and PA and would allow us to constrain the binary orbit to a much
higher degree. The wavelength-averaged closure phase measurements are
presented in Fig. \ref{FIG:VAR_CP}. All bar one of the closure phase
measurements indicate, or are consistent with, a symmetric
environment. Therefore, we do not definitively detect the binary
companion. To investigate the geometrical constraints supplied by the
closure phase observations, we fit the observations with two geometric
models: that of a binary and the skewed ring model of
\citet{Monnier2006}. The binary model has only two free parameters,
the binary PA and separation. The two stellar components are treated
as unresolved uniform discs and the binary flux ratio was set to 3.5,
as discussed in W12. The parameters of the skewed ring model are: the
ring radius and elongation, the ring flux, the ring PA, the degree of
skewness and the PA of the skewed distribution. The properties such as
the ring radius, inclination and flux were set to those of the ring
found by W12 to represent the continuum emission of this
object. Therefore our skewed ring model also has 2 unknowns: the
degree of skewness and the PA of the skew.

\smallskip

It was found that neither model could be excluded. Ultimately, the
observations hint at an increase and subsequent decrease in the
asymmetry of the environment of HD~327083. If this represents a change
in the amount of asymmetry in the structure resolved in W12, its
degree of asymmetry must increase on a timescale of several
months. This appears unlikely given the size of the structure
(approximately 5~AU). In addition, we note that the CO bandhead
emission is located close to and interior to the continuum
emission. Therefore, if the continuum emission varied, perhaps as a
result of intermittent mass loss or accretion, variability in the CO
emission would be expected. Since the CO emission is not variable, it
appears that the continuum emission originates in a stable
structure. Consequently, we attribute our tentative detection of a
change in closure phase to motion of the binary system. A conclusive
confirmation of the interferometric binary detection may require model
independent imaging with, for example, the PIONIER instrument
\citep[e.g. as demonstrated by][]{Blind2011}.

\section{Discussion}

\label{disc}

This paper presents high spectral resolution (25 \&
6~$\mathrm{km\,s^{-1}}$) AMBER and CRIRES observations of the binary
supergiant B[e] star HD~327083. The combination of information on AU
sized scales provided by the AMBER data and the velocity resolved
information provided by the CRIRES data offers new insights into the
close environment of this object. In particular, we show that the data
can be reproduced well with a Keplerian disc model. On the contrary,
the scenario of an equatorial outflow appears unlikely since such a
model cannot reproduce the observations.

\smallskip

These observations, and the disc model that best fits the CRIRES
spectrum, confirm several previous suggestions that were based on
spectrally unresolved interferometric observations. In W12 it was
remarked that there was no clear visibility signature across the CO
overtone emission. On this basis, it was suggested that the CO
emission originates from a region with the same elongation, and thus
inclination, as the near-IR continuum. Furthermore, in W12 it was
suggested that the CO material is located interior to the continuum
emission. The geometry of the CO bandhead emission predicted by the
spectral fit to the CRIRES spectrum is consistent with both these
hypotheses. The near-IR continuum itself is thought to be dominated by
thermal emission of warm dust. Therefore, the location of the CO emitting
region is consistent with the presence of a gaseous disc interior to the near-IR
continuum emitting region (which has an inner radius of
approximately 5~AU). This supports the association of the continuum
emitting region with the inner rim of a dusty disc.

\smallskip

The most compelling support for the disc scenario is provided by the
kinematic element of the modelling. The excellent agreement between
the model spectral profile and the data proves that the kinematics of
the emitting material are entirely consistent with Keplerian motion.
Finally, we note that the AMBER differential phase signature provides
a third argument in terms of photocentric displacement on scales of
0.15~mas. The observed signature is found to be fully consistent with
the predictions of the best fitting model. The combination of high
spatial and spectral precision provides stringent constraints on the
kinematics of line emission regions. Therefore, the excellent
agreement between the data and disc model in both the spatial and
spectral domains is compelling evidence that the CO emission
originates in a Keplerian disc.

\smallskip

As far as the nature of this disc is concerned, we observe the
following. We note that the relatively low inner temperature of the
disc suggests there is an absence of hot material close to the central
star. Consequently, this is consistent with the presence of an inner
gap (see Fig. \ref{FIG:model_comp}). The derived inner radius of the
CO emitting disc is $\sim$3~AU, which is larger than the (somewhat
uncertain) binary separation of $a\sim$1.7~AU (M2003). In other words,
the model presented here confirms the suggestion of W12 that the
HD~327083 system is surrounded by a circumbinary disc. The fact that
the gaseous disc does not extend closer to the central star, where its
inner rim temperature would be $\sim$5000~K, provides further insight
into the mass loss of this system. Specifically, the classification of
HD~327083 as an interacting $\beta$ Lyrae type binary (M2003) can most
likely be discounted, because of the absence of a circumstellar
accretion disc.  Also, the sgB[e] model appears unfitting for this
system \citep[HD~327083 is listed as a sgB$\mathrm{[e]}$ candidate
by][]{Kraus2009}. The CO bandhead emission of such objects is expected
to trace an equatorial outflow. Since we show that this model does not
reproduce the data well, it appears unlikely that this object can be
described by the standard sgB[e] picture. This supports the hypothesis
that the B[e] behaviour of HD~327083 is related to its binarity.

\smallskip

This conclusion raises the question of whether there are any other
systems similar to HD~327083. The CO emission contains two clues
regarding the origin of the material surrounding HD~327083. First, we
reiterate that the material is confined to a circumbinary
disc. Secondly, we note that the gaseous circumbinary disc exhibits no
evidence of variability. CO emission traces hot dense gas in the
inner-most regions of discs and is sensitive to the surface density of
the emitting region. Therefore, it is likely that the CO emission of
HD~327083 would be sensitive to varying mass injection into the
circumbinary disc. Consequently, the lack of variability suggests that
the circumbinary disc is a relatively stable structure.

\smallskip

Considering these points, the binary HD~327083 appears very similar to
the binary A[e] star HD~62623 recently studied by
\citet{Millour2011}. HD~62623 is also a binary and \citet{Millour2011}
demonstrate that a circumbinary disc surrounds the system. To explain
the characteristics of this system, \citet{Plets1995} proposed that
angular momentum transfer occurs through the L2 Lagrangian point of
the system. As a result, mass lost from the primary supergiant by its
radiatively driven wind is deposited into a stable dense circumbinary
disc \citep[which is consistent with the observations
of][]{Millour2011}. If this scenario applies to HD~327083, the mass
injection into the disc would be less dependent on orbital phase than
the Roche Lobe overflow at periastron suggested by M2003. Therefore,
this scenario could explain the lack of variability in the CO
emission. Furthermore, this scenario also explains the presence of a
circumbinary disc. Consequently, the environment of HD~327083 is
entirely consistent with this scenario. As a result, HD~327083 could
belong to the class of massive binaries which feature circumbinary
discs thought to be due to the channeling of the primary's wind
through the L2 point of the system \citep[currently exemplified by HD
62623 and possibly also $\upsilon$ Sgr,
see][]{Millour2011,Bonneau2011}.

\smallskip

As a final remark, we note that this paper demonstrates the diagnostic
potential of high spectral resolution observations of CO bandhead
emission coupled with information on small angular scales as provided
by VLTI observations. This approach offers a means to directly probe
the circumstellar kinematics of sgB[e] stars. Recent observations
have found additional evidence for Keplerian structures around other
sgB[e] stars, contrary to expectations based on the 'dual outflow'
sgB[e] model \citep[][]{Aret2012}. Therefore, it is clear we have yet
to arrive at a complete understanding of these objects. To
systematically study the circumstellar environments of sgB[e] stars,
we have conducted a CRIRES survey of Galactic sgB[e] and B[e]
stars. The CRIRES observations presented here were drawn from this
survey and demonstrate the quality and potential of the data. Analysis
of the entire sample will be presented in a forthcoming paper
(Muratore et al. in prep.) and follow-up VLTI observations are
planned.

\section{Conclusion}

\label{conc}

This paper presents high spectral resolution (25 \&
6~$\mathrm{km\,s^{-1}}$) AMBER and CRIRES
observations of the supergiant B[e] star HD~327083.  The
observations spectrally resolve the CO bandhead emission of this
object for the first time and provide new insights into its immediate
environment. Here we list the key results.

\begin{itemize}

\item[--]{We find no evidence that the CO bandhead emission of
  HD~327083 is variable. Since the observations span the majority of
  the estimated period of the binary system, this suggests that the
  structure of the CO emitting region is not strongly dependent on
  orbital phase.}

\item[--]{The CRIRES observations of HD~327083's CO $\mathrm{1^{st}}$
  overtone bandhead emission are fit to an excellent degree by a model
  of a Keplerian rotating disc.}

\item[--]{A differential phase signature corresponding to a
  photo-centre displacement of $\sim$0.15~mas is observed. This is
  well reproduced by the same model that fits the high spectral
  resolution CRIRES spectrum. The agreement between the model and
  observations in both the spatial and spectral domains is compelling
  evidence that the CO emission of HD~327083 originates in a Keplerian
  disc.}

\item[--]{The inner edge of the best fitting CO disc is approximately
  3$\pm$0.3~AU. This places it interior to the previously detected
  continuum emitting disc (which has an inner radius of approximately
  5~AU), but is sufficiently large for the CO emitting disc to be
  circumbinary.}

\item[--]{We show that a model of an equatorial outflow cannot
  reproduce the data as well as the Keplerian disc model. This
  suggests that the standard sgB[e] scenario is not applicable to
  HD~327083, which supports the hypothesis that the B[e] behaviour of
  HD~327083 is due to binarity.}

\end{itemize}

\begin{acknowledgements}
  An anonymous referee is thanked for his/her constructive
  comments. HEW acknowledges the financial support of the MPIfR in
  Bonn. JDI is grateful for a studentship from the Science and
  Technology Facilities Council of the UK. A. Liermann is thanked for
  a careful reading of the manuscript. The NSO/Kitt Peak FTS data used
  here were produced by NSF/NOAO. This paper makes use of the high
  spectral resolution $K$-band library of standard spectra provided by
  the GEMINI Observatory, which is operated by the Association of
  Universities for Research in Astronomy, Inc., under a cooperative
  agreement with the NSF on behalf of the international Gemini
  partnership of Argentina, Australia, Brazil, Canada, Chile, the
  United Kingdom, and the United States of America. Finally, this
  research has made use of the \texttt{AMBER data reduction package}
  of the Jean-Marie Mariotti Center\footnote{Available at
    http://www.jmmc.fr/amberdrs}.

\end{acknowledgements}

\bibliographystyle{aa} 
\bibliography{bib}

\end{document}